\documentclass{icrc29}
\usepackage{graphicx,amssymb,amsmath,times}
\begin{document}

\title[Large-scale anisotropy of EGRET gamma ray sources]{Large-scale anisotropy of EGRET gamma ray sources}
\author[Luis Anchordoqui et al.] {Luis Anchordoqui$^a$, Thomas McCauley$^b$, Thomas  Paul$^a$,
       Olaf Reimer$^c$,  Diego F. Torres$^d$\\
        (a) Department of Physics, Northeastern University, Boston, MA 02115 USA\\
        (b) Institute for Nuclear and Particle Astrophysics, Lawrence Berkeley National Laboratory, Berkeley,             CA 94720 USA\\
        (c) W.W.Hansen Experimental Physics Laboratory, Stanford University, Stanford, CA 94305 USA\\
        (d) Lawrence Livermore National Laboratory, Livermore, CA 94551 USA\\
        }
\presenter{Presenter: Thomas Paul (tom.paul@cern.ch), \  usa-mccauley-T-abs1-og22-oral}

\maketitle

\begin{abstract}

In the course of its operation, the EGRET experiment detected high-energy
$\gamma$-ray sources at energies above 100 MeV over the whole sky. In this
communication, we search for large-scale anisotropy patterns among the 
catalogued EGRET sources using an expansion in
spherical harmonics, accounting for EGRET's highly non-uniform exposure. 
We find significant excess in the quadrupole and octopole moments.  This is consistent
with the hypothesis that, in addition to the galactic plane, a second 
mid-latitude ($5^{\circ} < |b| < 30^{\circ}$) population, perhaps associated 
with the Gould belt, contributes
to the $\gamma$-ray flux above 100~MeV.

\end{abstract}

\section{Introduction}

Our rudimentary understanding of the GeV $\gamma$-ray sky was 
greatly advanced in 1991 with the launch of the Energetic Gamma Ray 
Experiment Telescope (EGRET) on board  the Compton Gamma Ray Observatory 
(CGRO). The science returns from EGRET observations have exceeded pre-launch 
expectations, increasing the number of known 
$\gamma$-ray sources from 1--2 dozen to the 
271 listed in the third EGRET (3EG) catalog~\cite{Hartman:1999fc}.  
However, of this 
multitude of sources, only about half  have been {\em definitively} associated with 
known astrophysical objects.  
One reason so few sources have been uniquely identified
is due to a typical location uncertainty of up to $1^\circ$ for an EGRET
detected gamma-ray source, an area 
typically containing several potential candidate astrophysical objects.
Consequently most of the $\gamma$-ray sky, as we currently 
understand it, consists of unidentified sources. 

The EGRET sources appear to arise from several distinct populations. There is an almost 
isotropic distribution of $\sim 100$ low/high variability objects, nearly all of which have now been 
plausibly associated with flaring blazars, a radio bright sub-class of active 
galactic nuclei~\cite{vonMontigny:1995yz,Mattox}. The remaining sources are believed to 
be within the Milky Way, and there is some speculation that they may comprise two 
distinct classes.  The first class consists of bright sources with generally hard spectra 
with galactic latitude $ |b| < 5^\circ$ of the Galactic Plane, and are thought to be bright objects
up to aproximately 10 kpc away. There is some evidence of a second population at mid-Galactic latitudes
($5^{\circ} < |b| < 30^{\circ}$) consisting of fainter objects with generally steeper
spectra. It has been hypothesized that this population is associated with the 
Gould belt~\cite{Gehrels:2000gp,Grenier}, an asymmetric
structure in a great circle on the sky tilted $20^\circ$ to the Galactic Plane.  The Gould belt
comprises massive O and B stars as well as clusters of molecular clouds and expanding interstellar
gas at distance of around 200 pc from earth.  The observation of spatial 
coincidence~\cite{Sturner:1994jc,Romero:1999tk,Torres:2002af} of low-latitude 
($|b| < 10^\circ$) EGRET sources 
with massive OB stars and supernova remnants provides additional indication that the 
mid-latitude sources originate in the Gould belt.

\begin{figure}[!thb]
\begin{minipage}[t]{0.48\textwidth}
\mbox{}\\
\centerline{\includegraphics[width=\textwidth]{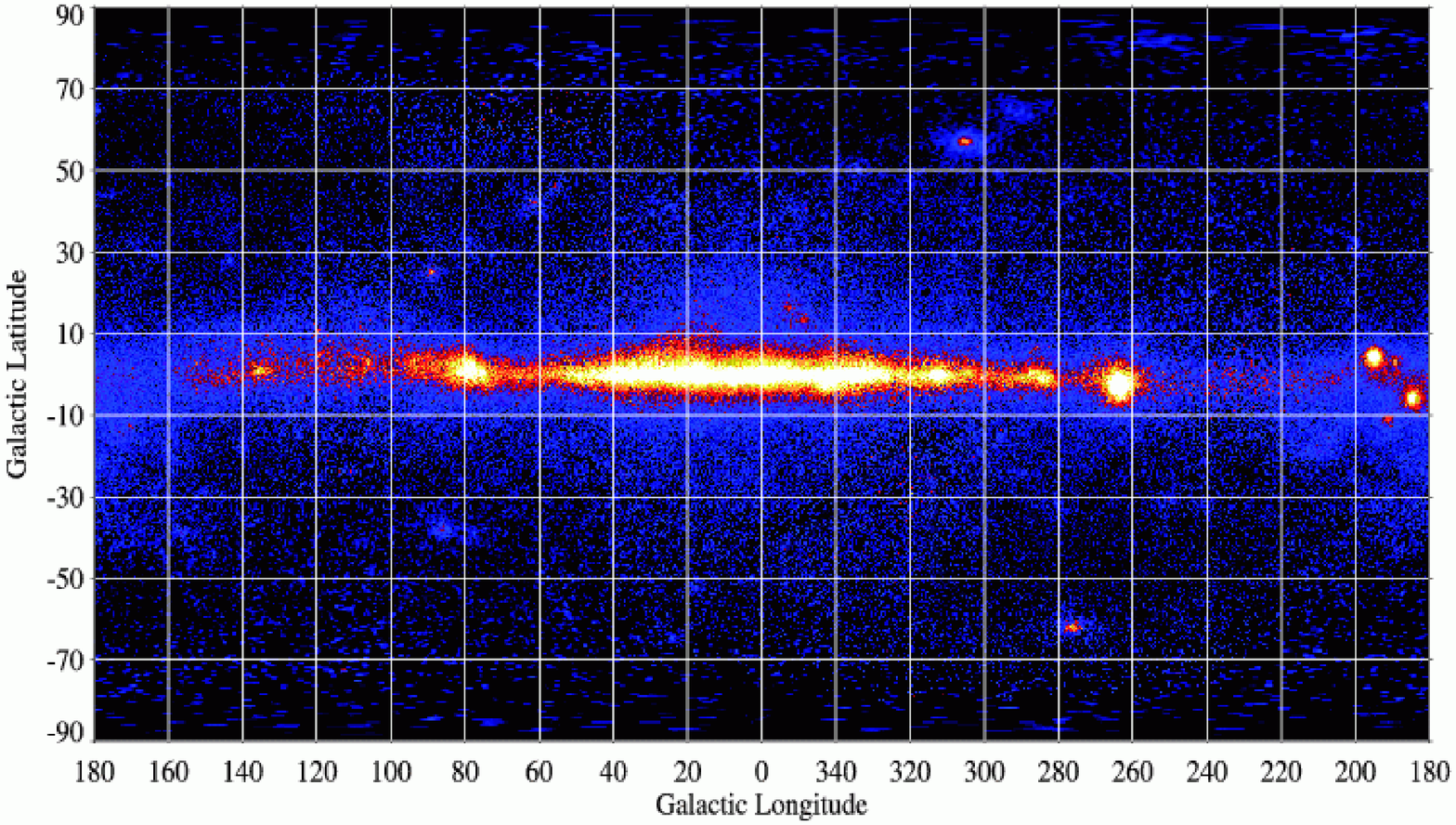}}
\caption{\label{egret_intensity} Intensity map of EGRET $\gamma$-ray sources with energies above 100~MeV.}
\end{minipage}
\hfill
\begin{minipage}[t]{0.50\textwidth}
\mbox{}\\
\centerline{\includegraphics[width=\textwidth]{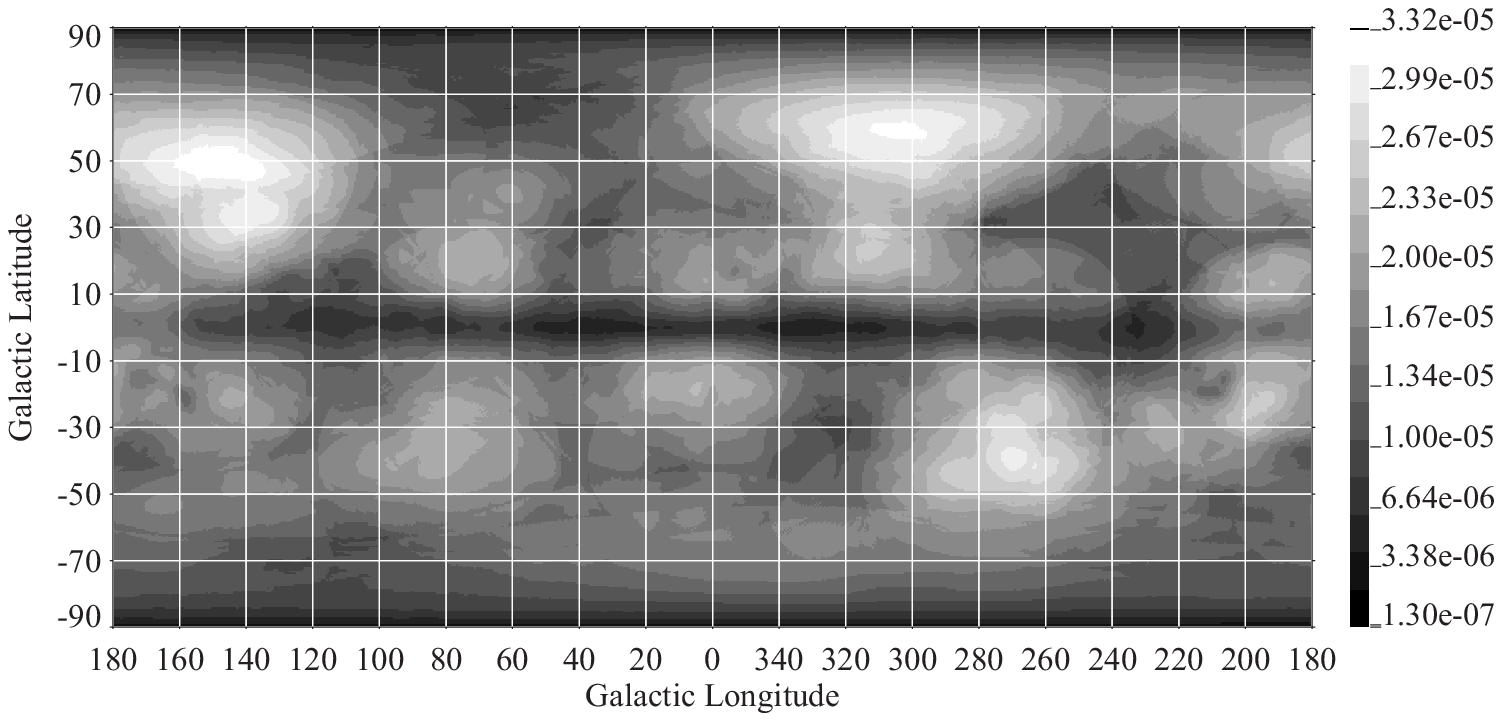}}
\caption{\label{egret_exposure} Two-dimensional EGRET detectability map for 
$\gamma$-ray sources (in units of cm$^2$ s).}
\end{minipage}
\end{figure}

Figure~\ref{egret_intensity} shows the intensity map of EGRET sources.  
Though the correlation with the Galactic Plane is obvious, the hypthosized
Gould belt population is not easy to extract using an eyeball fit.
In this communication, we apply a power spectrum analysis~\cite{Peebles} in an attempt
to tease this component out of the bright Galactic Plane sources.

\section{EGRET Source Detectability}

For $E_\gamma > 100$~MeV, the significance for a detection of an isolated point source with EGRET 
scales as
\begin{equation}
\omega_j \propto F \sqrt{{\cal E}/{\cal B}} \,\,,
\end{equation}
where $F$ is the flux, ${\cal E}$ is the exposure, and ${\cal B}$ is the intensity of the diffuse 
$\gamma$-ray emission at the region of the source~\cite{Mattox:2}. Following~\cite{Reimer:2001ba}, 
we derive the detectability map considering these three ingredients in turn.
First, the populations used in the Monte Carlo simulations are prepared such that
they reproduce the observed distribution of the number of sources per flux bin,
globally as well as locally.  Second, it is necessary to account for the highly 
non-uniform exposure of the instrument.  Thirdly, the diffuse $\gamma$-ray background
is compensated using the EGRET diffuse model, corresponding to the commonly believed cosmic ray/gas
distribution coupling (dynamic balance). 

With these considerations in mind, we construct the EGRET 
detectability map by 
determining 95\% CL limits for a grid on the sky.  We determine the limits using the
so-called P1234 time span~\cite{Hartman:1999fc},  {\it i.e.}, we request
that the maximally significant measurement of the EGRET source has been
obtained in the P1234 joint viewing period. This tends to eliminate flaring
sources from our sample. The map is shown in Figure~\ref{egret_exposure}.
To obtain the detectability weights we normalize each angular bin to unity.

\section{Angular Power Spectrum}

The spherical harmonics form a useful set for 
expansions of the intensity over the celestial sphere. For a uniform sky coverage, 
\begin{equation}
I ({\bf n})= \sum_{\ell = 0}^\infty \,\, 
\sum_{m = -\ell}^\ell\,\, a_{\ell m}\,  Y_{\ell m}({\bf n})\,\, ,
\label{CP20}
\end{equation}
where ${\bf n}$ is a unit vector that denotes the direction in the sky of each source in the survey.\footnote{Throughout this work 
we use real-valued spherical harmonics, which are obtained from the complex ones by substituting, 
$e^{i\,m\,\phi} \rightarrow \sqrt{2} \sin (m\phi)$, if $m<0$, $e^{i\,m\,\phi} \rightarrow \sqrt{2} 
\cos (m\phi)$, if $m>0$, and $e^{i\,m\,\phi} \rightarrow 1$ if $m=0.$}
To account for the nonuniform EGRET exposure, we define a weighted intensity
\begin{equation}
I({\bf{n}}) = \frac{1}{\cal N}\,\,\sum_{j = 1}^N  \frac{1}{\omega_j} \,\, \delta ({\bf n}, {\bf n}_j) \,\,, 
\label{I}
\end{equation}
where $\omega_j$ is the relative detectability at sky direction ${\bf n}_j$, ${\cal N}$ 
is the sum of the weights $\omega_j^{-1}$, and $N$ is the number of sources. 
Since the eigenvalues of the $Y_{\ell m}$ expansion are 
uniquely defined
\begin{equation}
a_{\ell m} = \int  I ({\bf n}) \,\, Y_{\ell m} ({\bf n}) \,\, d^2{\bf n}\,\,,
\label{aintegral}
\end{equation}
the replacement of Eq.~(\ref{I}) into Eq.~(\ref{aintegral}) leads to the explicit form 
of the coefficients for our set of $N=124$ EGRET sources 
\begin{equation}
a_{\ell m}=\frac{1}{{\cal N}}\sum_{j=1}^N \, \frac{1}{\omega_j} \,\, Y_{\ell m}({\bf n}_j) \,.
\label{CP21}
\end{equation}
The coordinate independent angular power spectrum of fluctuations, 
\begin{equation}
C(\ell) = \frac{1}{(2 \ell +1)}\,\, \sum_{m=-\ell}^\ell a_{\ell m}^2\,\,,
\label{CP23}
\end{equation}
provides a gross summary of the features present in the celestial distribution together with the 
characteristic angular scales. 
The power in mode $\ell$ is sensitive to variation over 
angular scales of  $\ell^{-1}$ radians.

If the steady EGRET sources in fact originate from two overlapping great circles 
associated with the Galactic Plane and the Gould belt, one would expect
a strong signal in the quadrupole ($\ell=2$) moment describing the main 
component, and a fainter signal in the octopole ($\ell=4$) moment indicating
the spread from a perfect quadrupole.  Note that the Gould belt is tilted 20$^\circ$
relative to the Galactic Plane, so one would expect its presence to be
mainifest in the octopole term which is sensitive to angular scales of about 15$^\circ$.

\begin{figure}[!thb]
\begin{minipage}[t]{0.48\textwidth}
\mbox{}\\
\centerline{\includegraphics[width=\textwidth]{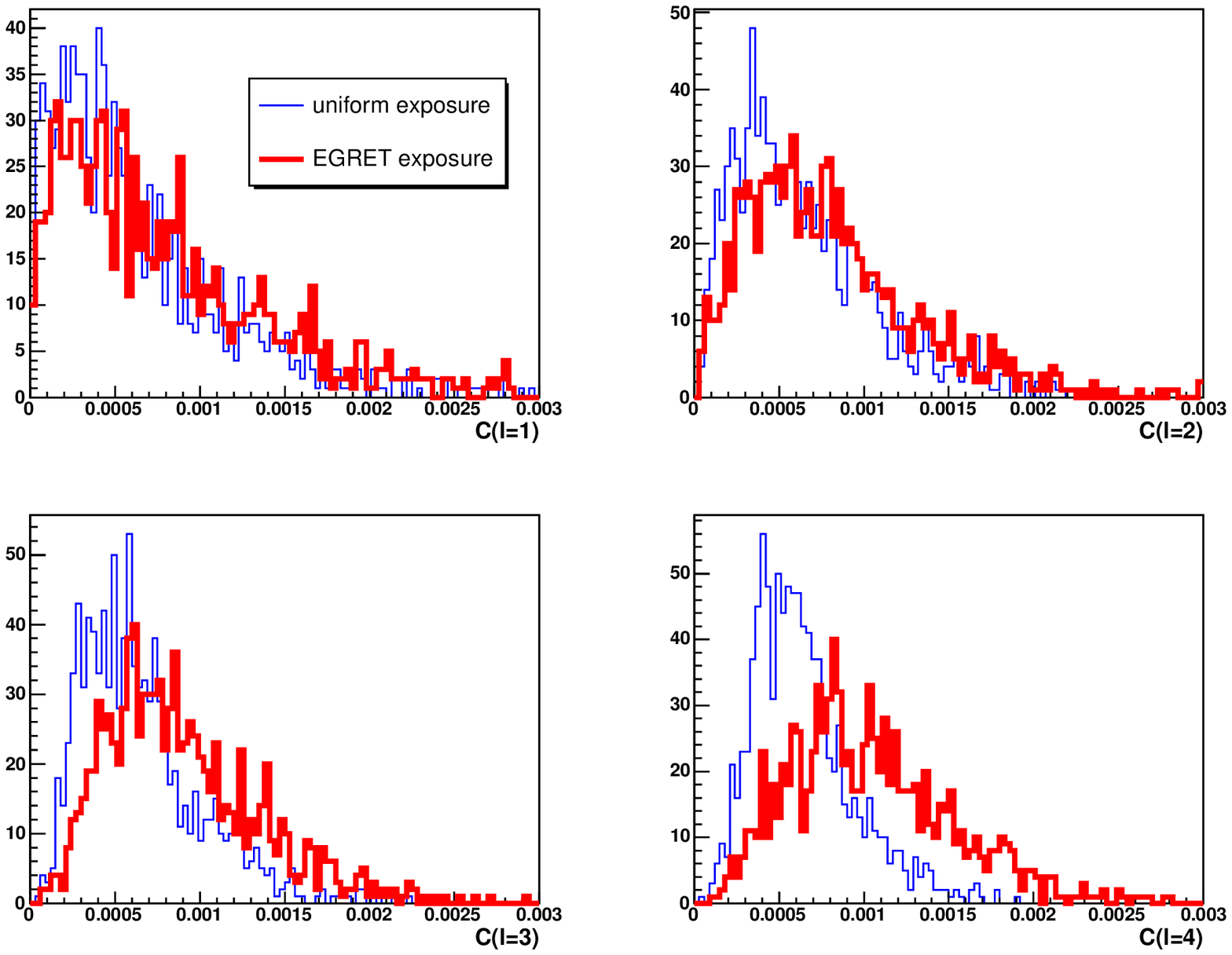}}
\caption{\label{Cl_hist} Comparison of the distributions of 
angular power spectra for isotropically distributed sources observed
with uniform exposure and with the EGRET exposure.}
\end{minipage}
\hfill
\begin{minipage}[t]{0.48\textwidth}
\mbox{}\\
\centerline{\includegraphics[width=\textwidth]{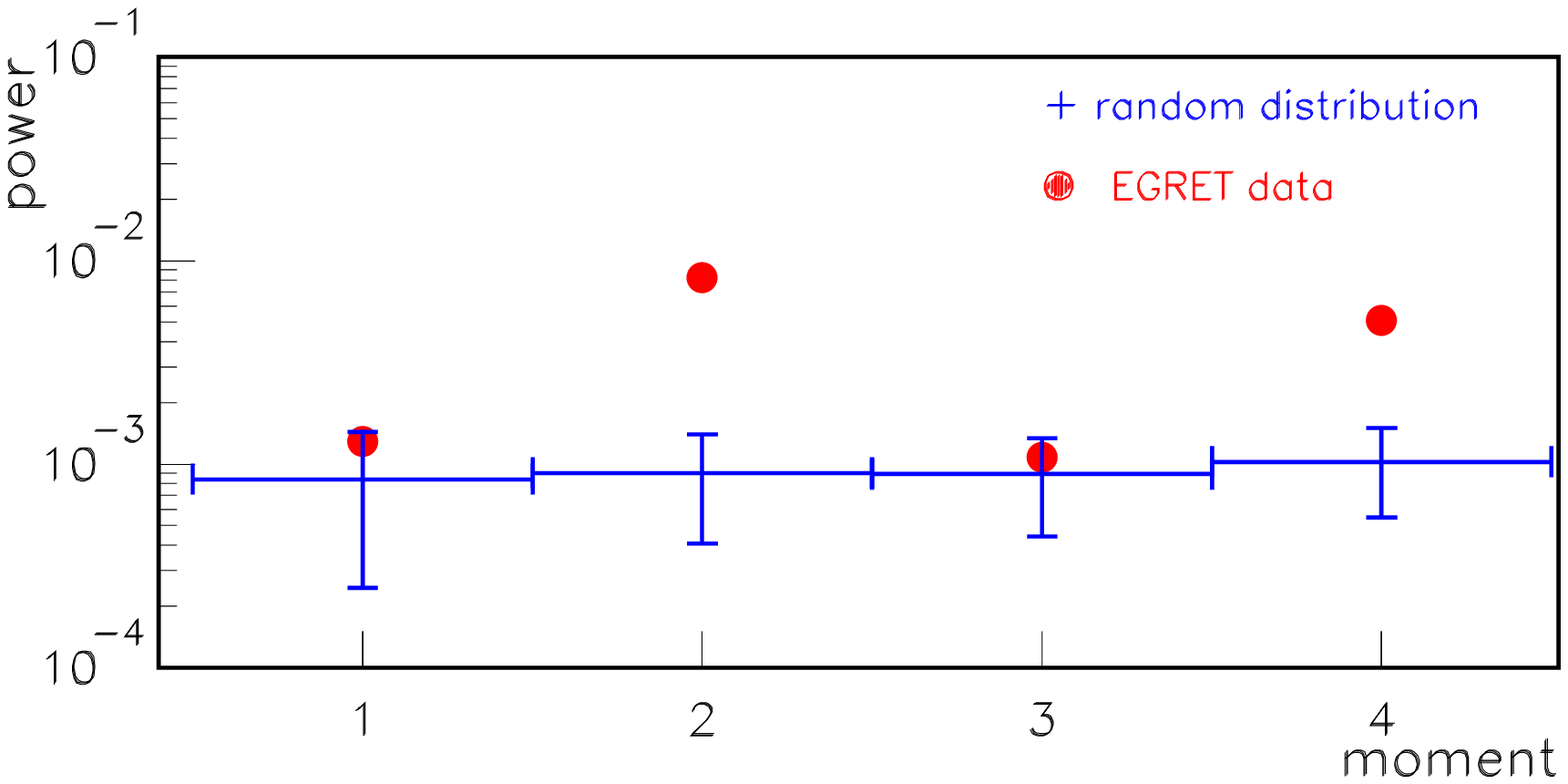}}
\caption{\label{Cl_egret} The angular power spectrum of the EGRET four-year observations is 
indicated by circles. 
The lines with error bars show the RMS of $\overline C_\ell$. 
The RMS values are obtained 
from 1000 sets of Monte Carlo simulations of 124 events each, including corrections for $\omega_j$.}
\end{minipage}
\end{figure}

For $\ell \geq 1,$ the expectation for an isotropic distribution observed with
 uniform exposure is 
$\overline{\cal C}_\ell = (4 \pi N)^{-1},$ and the RMS fluctuations are~\cite{Anchordoqui:2003bx}
\begin{equation}
\overline {\Delta {\cal C}}_\ell = \sqrt{2/(2 \ell +1)} \,\, \overline{\cal C}_\ell \,\,.
\end{equation}
Figure~\ref{Cl_hist} shows a comparison of the distributions of 
angular power spectra for isotropically distributed sources observed
with uniform exposure ($\overline {\cal C}_\ell$) and with the EGRET exposure ($\overline C_l$).  
It can be seen that for the four multipole moments of interest, the spread of the 
distributions corresponding to uniform and non-uniform exposures are 
similar.

Figure~\ref{Cl_egret} compares the observed angular power spectrum of EGRET to 
the spectrum expected for an isotropic distribution.  Clear excesses are seen for
the quadrupole and octopole moments. 
 The probability of the excess for either moment to be consistent with isotropy is estimated 
(from a fit to the distributions shown in Fig.~\ref{Cl_hist})
to be less than ${\cal O}(10^{-10}).$ This supports the hypothesis that,
in addition to the Galactic Plane, a mid-latitude population of sources, 
associated with the Gould belt, contributes
to the observed $\gamma$-ray flux above 100~MeV.

\section*{Acknowledgements}
The work of LA and TP has been partially supported by the US NSF, Grant No. PHY-0457004.
TM acknowledges the support of the US DOE under contract number
DE-AC-76SF00098. OR acknowledges support by DLR QV0002. 
The work of DFT was performed under the auspices of the US DoE by University of California's LLNL 
under contract No. W-7405-Eng-48.


\begin{thebibliography}{99}

\bibitem{Hartman:1999fc}
  R.~C.~Hartman {\it et al.}  [EGRET Collaboration],
  Astrophys.\ J.\ Suppl.\  {\bf 123}, 79 (1999).

\bibitem{vonMontigny:1995yz}
  C.~von Montigny {\it et al.},
  Astrophys.\ J.\  {\bf 440}, 525 (1995).

\bibitem{Mattox} 
  J. R. Mattox, R. C. Hartman and O. Reimer,
  Astrophys.\ J.\ Suppl. {\bf 135}, 155 (2001).

\bibitem{Gehrels:2000gp}
  N.~Gehrels, D.~J.~Macomb, D.~L.~Bertsch, D.~J.~Thompson and R.~C.~Hartman,
  Nature {\bf 404}, 363 (2000).


\bibitem{Grenier} 
   I. A. Grenier
   Adv. Space Res. {\bf 15}, 73 (1993).

\bibitem{Sturner:1994jc}
  S.~J.~Sturner and C.~D.~Dermer,
  Astron.\  Astrophys.\ {\bf 293} L17 (1995). 

\bibitem{Romero:1999tk}
  G.~E.~Romero, P.~Benaglia and D.~F.~Torres,
  Astron.\ Astrophys.\  {\bf 348}, 868 (1999).


\bibitem{Torres:2002af}
  D.~F.~Torres, G.~E.~Romero, T.~M.~Dame, J.~A.~Combi and Y.~M.~Butt,
  Phys.\ Rept.\  {\bf 382}, 303 (2003).



\bibitem{Peebles} 
  P. J. E. Peebles, 
  Astrophys. J. {\bf 185}, 413 (1973).



\bibitem{Mattox:2} 
  J. R. Mattox {\it et al.},
  Astrophys.\ J.\ {\bf 461}, 396 (1996).





\bibitem{Reimer:2001ba}
  O.~Reimer,
  in {\em The Nature of Unidentified Galactic
  High-Energy Gamma-Ray Sources,} ISBN 1-4020-0010-3 (Astrophysics and
  Space Science Library), Kluwer, 2000, p.17-34








\bibitem{Anchordoqui:2003bx}
  L.~A.~Anchordoqui, C.~Hojvat, T.~P.~McCauley, T.~C.~Paul, S.~Reucroft, J.~D.~Swain and A.~Widom,
  Phys.\ Rev.\ D {\bf 68}, 083004 (2003).


\end{thebibliography}
\end{document}